\documentclass[prd,opreprint,twocolumn,superscriptaddress,showpacs,floatfix,amsmath,amssymb,floatfix,nofootinbib]{revtex4}
\pdfoutput=1
\usepackage{dcolumn}
\usepackage{bm}
\usepackage{epsfig}
\usepackage{amsmath}
\usepackage{subfigure}
\usepackage{amsfonts}
\usepackage{amssymb}
\usepackage{graphicx}
\usepackage{pstricks}
\usepackage{array}
\usepackage{hyperref}
\usepackage{multirow}
\usepackage{caption}
\usepackage{float}

\DeclareSymbolFont{AMSa}{U}{msa}{m}{n}
\DeclareSymbolFont{AMSb}{U}{msb}{m}{n}
\let\Box\relax
\DeclareMathSymbol{\Box}{\mathord}{AMSa}{"03}

\newcommand{\be}{\begin{equation}}
\newcommand{\ee}{\end{equation}}
\newcommand{\bea}{\begin{eqnarray}}
\newcommand{\eea}{\end{eqnarray}}

\newcommand{\eq}[1]{\begin{equation}\begin{split} #1 \end{split}\end{equation}}

\newcommand{\ds}{\displaystyle}

\def \abs#1{\left\vert#1\right\vert}

\usepackage{pslatex}
\usepackage[T1]{fontenc}
\usepackage{graphicx}
\usepackage{epsfig}

\usepackage{calc}
\usepackage{ifthen}
\usepackage{subfigure}

\newcommand{\lh}{\lambda_h}
\newcommand{\ls}{\lambda_s}

\begin{document}
\title{From inflation to cosmological electroweak phase transition with a complex scalar singlet}
\author{Wei Cheng}
\affiliation{Department of Physics, Chongqing University, Chongqing 401331, China}

\author{Ligong Bian}
\email{lgbycl@cqu.edu.cn}
\affiliation{Department of Physics, Chongqing University, Chongqing 401331, China}
\affiliation{Department of Physics, Chung-Ang University, Seoul 06974, Korea}

\begin{abstract}
In this work, we investigate the possibility to realize inflation and strong first order electroweak phase transitions(SFOEWPT) together with the relic density explanation. We studied the usual Higgs-portal real singlet dark matter model and the global U(1) breaking complex singlet model with the pseudoscalar being dark matter candidate. We focus on higher dark matter parameter regions due to the lower dark matter masses region are almost excluded by the current direct detection experiments except that of the so-called ${\it Higgs~ funnel ~regime}$ with $m_{DM}\sim m_h/2$. In both models, the SFOEWPT can occur through two-step where the magnitude of the Higgs-singlet quartic couplings can account for the Higgs or singlet inflation. The usual Higgs-portal real singlet dark matter model can not address the correct relic density together with the 
explanation of inflation and SFOEWPT. With the complex singlet model, the correct relic density can also be obtained simultaneously when 
the mixing angle $0.1<\theta<0.2$ and the dark matter mass is
$\sim800$ GeV. 
 
\end{abstract}
\maketitle
\preprint{}


\section{Introduction}

With the observation of the 126 GeV standard model (SM) Higgs, it seems the SM of the particle physics is complete. Meanwhile, the searching for new physics at LHC is continuing. On the other hand, the absent of dark matter in the model make it unsatisfactory. The baryon asymmetry of the Universe has puzzled the particle and cosmology physicists for decades. The two shortcomes of the SM motivated a lot of papers from particle and cosmology physicists as well as astrophysicists, e.g.,~\cite{Liu:2017gfg,Gu:2017rzz,Chao:2017vrq,Chao:2014ina,Ghorbani:2017jls,AbdusSalam:2013eya,
Fairbairn:2013uta,Ahriche:2013zwa,Cline:2013bln,Falkowski:2012fb,Gil:2012ya,Borah:2012pu,Ahriche:2012ei,Chowdhury:2011ga,Carena:2011jy,Chung:2011it,
Dimopoulos:1990ai}. At the same time, the horizon and homogeneity problems of the universe has been addressed by the inflation very well~\cite{Guth:1980zm}. Recently, several attempts have been motivated to explain the dark matter and inflations, see.~\cite{Lerner:2009xg,Tenkanen:2016idg,Choubey:2017hsq,Hamada:2017sga,Inomata:2017bwi,Cado:2016kdp,Ballesteros:2016euj}.

The mechanism of  Electroweak brayogenesis can explain the baryon asymmetry of the Universe with the cosmological phase transition happen at electroweak scale where the extend of the Higgs sector is necessary, which can be detected through the modification of the triple higgs couplings at the collider, see.~\cite{Arkani-Hamed:2015vfh}, that make it very attractive. For a recent review of Electroweak brayogenesis, see Ref.~\cite{Morrissey:2012db}.
The singlet scalar extended SM has been employed to study the electroweak phase transition(EWPT) extensively,  see Ref.~\cite{Espinosa:2011ax,Profumo:2014opa} for gauge singlet real scalar case, and Ref.\cite{Jiang:2015cwa,Chiang:2017nmu} for the complex gauge singlet scalar cases.

When the gauge singlet real scalar carry some discrete symmetry( e.g., the $Z_2$ symmetry), it can serve as the Weakly Interacting Massive Particle (WIMP) dark matter candidate. As for the 
complex singlet, both two configurations can serve as dark matter candidates when $U(1)$ symmetry is preserved~\cite{Barger:2008jx,McDonald:1993ex}.
The Higgs portal singlet scalar dark matter has been studied extensively due to it's simplicity and economy, see \cite{Cline:2012hg,Cline:2013gha} for recent studies. Up to now,
with the accumulating of direct detection experimental data, the parameter spaces are severely constrained, dark matter allowed region is pushed up to around TeV for real and complex dark matter, see Ref.~\cite{Wu:2016mbe,Escudero:2016gzx,Han:2015hda} except the $Higg ~funnel ~regime$ $m_{DM}\sim m_h/2$.

The Electroweak brayogenesis (EWBG) in the gauge singlet model is impossible when an additional CP violation phase is absent. 
An extra dim-6 operator at
zero temperature is useful to account for the chiral asymmetry during the process of EWPT for real singlet~\cite{Vaskonen:2016yiu}
as well as the complex singlet model case as been studied in \cite{Jiang:2015cwa} recently, where the dark matter relic abundance can be explained also.
In the complex singlet model, see \cite{Barger:2008jx,Gonderinger:2012rd,Barger:2010yn,Gonderinger:2009jp}, 
the global $U(1)$ symmetry can be broken and make the pseudo-scalar being the WIMP dark matter candidate.
And the cancellation of direct detection happens due to the mixing of the real part of the singlet and the SM-like Higgs, as highlight in Ref.~\cite{Gross:2017dan}, can make more WIMP parameter spaces being survived with the accumulation data of dark matter direct detection experiments.

Motivated by above arguments, we will investigate the possibility to explain inflation, EWPT, together with dark matter relic abundance with the complex singlet model.
The model is described in Sec.~\ref{sec:mod}, where the gauge singlet real scalar model can be obtained when the global U(1) being reduced to $Z_2$ symmetry with the U(1) breaking terms absent. The ingredients for cosmic inflation, SFOEWPT, and dark matter in the model are given in Sec.~\ref{sec:ipd}.
The numerical results of the whole physical pictures of the three components are accomplished in Sec.~\ref{sec:res}. We conclude in Sec.~\ref{sec:condis}.

\section{The model}
\label{sec:mod}
\begin{figure}[t]
\begin{center}
\includegraphics[width=0.4\textwidth]{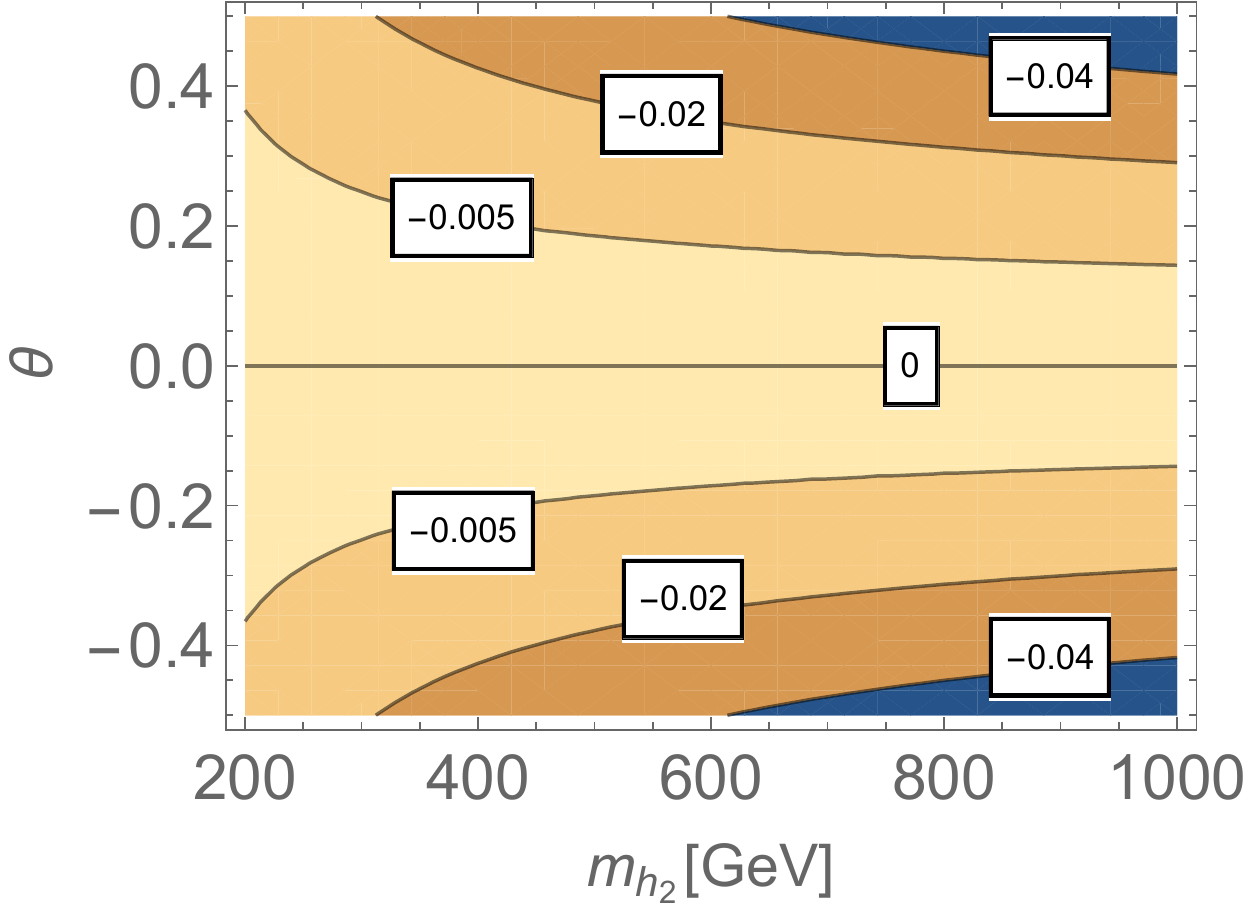}
\caption{$\Delta$T parameter in the parameter spaces of $\theta$ and heavy Higgs mass $m_{h_2}$. }
\label{fig:Tpar}
\end{center}
\end{figure}

In this work, we employ complex singlet scalars model with the tree level potential of the model being given by,
\begin{eqnarray}\label{tree}
V_0(H, \mathcal{S}) &=&  - \mu_h^2 |H|^2 + \lambda_h |H|^4 - \mu_s^2 |\mathcal{S}|^2 + \lambda_{hs}\abs{H}^2\abs{\mathcal{S}}^2\nonumber \\
&& + \lambda_s \abs{\mathcal{S}}^4 - (\frac{1}{2}\mu_b^2 \mathcal{S}^2 +h.c.)\;.
\end{eqnarray}
A real mass term $\mu_b$ is introduced to break the global U(1) $\mathcal{S}\rightarrow e^{i\alpha}\mathcal{S}$ symmetry, which make the imaginary parts of $\mathcal{S}$ serve as a dark matter candidate. After insert the scalar field configurations: $H^T = ( 0,~ h )/\sqrt 2~\rm{ and}~ \mathcal{S}=(s+\rm{I}~A)/\sqrt{2}$, we obtain,
 \begin{eqnarray}\label{tree2}
V_0(h, s,\chi) &=& \frac{ \lambda_h h^4}{4}+\frac{1}{4} \lambda_{hs} h^2 A^2-\frac{\mu _h^2 h^2}{2}+\frac{1}{4} \lambda_{hs}h^2 s^2+\frac{\lambda_s A^4}{4}\nonumber\\
&&-\frac{\mu_s^2 A ^2}{2}+\frac{\mu_{b}^2 A^2}{2}+\frac{\lambda_s s^4}{4}+\frac{1}{2} \lambda_s s^2 A^2-\frac{\mu_s^2 s^2}{2}\nonumber\\
&&-\frac{\mu_{b}^2  s^2}{2}.
 \end{eqnarray}
 The vacuum stability requires the tree-level potential should be bounded from below,
\begin{eqnarray}\label{stab}
\lambda_h > 0,~ \lambda_s> 0, ~ \lambda_{sh} >0 {~\rm or} ~ \lambda_{sh}> - 2\sqrt{\lambda_h\lambda_s}.
 \end{eqnarray}
Here, the stability conditions should be satisfied until the renormalization group(RG) scale being the Planck scale for safety of inflation and EWPT, which have been computed with the renormalization group equations list in sec.\ref{ap:B}.
The minimization conditions of the potential are,
\begin{eqnarray}
\frac{dV_0(h,s,A)}{dh}\big{|}_{h=v}=0,~\frac{dV_0(h,s,A)}{ds}\big{|}_{s=v_s}=0,
 \end{eqnarray}%
which give rise to $\mu_h^2=\lambda_h v^2+\lambda _{hs} v_s^2/2,\mu_s^2= -\mu_b^2+\lambda_{hs} v^2/2+\lambda_s v_s^2$. The mass matrix is then given by
\begin{eqnarray}
\mathcal{M}^2=
\bigg(\begin{array}{*{20}{c}}
{2{v^2}\lambda_h}&{vv_s\lambda_{hs}}\\
{vv_s\lambda_{hs}}&2{v_s^2\lambda_s}
\end{array}\bigg)\; .
\end{eqnarray}
In order to diagonalize the mass matrix, we introducing the rotation matrix $R= \left((\cos\theta ,\sin\theta ) ,( -\sin\theta ,\cos\theta ) \right)$ with $\tan2\theta  =  - {\lambda_{hs}vv_s}/{(\lambda_h{v^2} - \lambda_s v_s^2)}$ to relate the mass basis and field basis,
\begin{eqnarray}
\bigg(\begin{array}{*{20}{c}}
h_1\\
h_2
\end{array}\bigg)=
\bigg(\begin{array}{*{20}{c}}
\cos\theta&\sin\theta\\
-\sin\theta&\cos\theta
\end{array}\bigg)\bigg(\begin{array}{*{20}{c}}
h\\
s
\end{array}\bigg)
\end{eqnarray}
The mass squared eigenvalues are obtained as,
\begin{eqnarray}
m_{{h_1,h_2}}^2 =\lh v^2+\ls v_s^2\mp \frac{ \ls v_s^2 - \lh v^2}{\cos 2 \theta}\; .
\end{eqnarray}
Identify the $h_1$ being the 126~GeV SM-like Higgs boson, and requiring the $h_2$ is dominated by $s$ set $\sin\theta>1/\sqrt{2}$.
The quartic couplings can be expressed as functions of the Higgs masses, $v$, $v_{s}$ and the mixing angle $\theta$,
\begin{eqnarray}
\lambda_h&=&\frac{ \cos(2 \theta) \left(m_{h_1}^2-m_{h_2}^2\right)+m_{h_1}^2+m_{h_2}^2}{4 v^2},\\
\lambda_s&=&\frac{\cos(2 \theta ) \left(m_{h_2}^2-m_{h_1}^2\right)+m_{h_1}^2+m_{h_2}^2}{4 v_s^2},\\
\lambda_{hs}&=&\frac{tan(2\theta) \cos(2\theta)\left(m_{h_2}^2-m_{h_1}^2\right)}{2 v v_s}.
\end{eqnarray}
The mixing angle and the heavy Higgs masses are subjective to the bounds coming from the LHC Higgs data, which force the mixing angle $\theta$ to be larger than: $\abs{\cos\theta}\geq 0.84 $~\cite{Profumo:2014opa}. 
The mixing of the $h,s$ may leads to T parameter violation, which set stringent bounds on the mixing angle and the masses of the heavy Higgs. One can obtain the oblique parameter T with the formula of, 
\begin{eqnarray}
\nonumber
 \label{eq:Txsm1}
T & = & -\left(\frac{3}{16\pi s_W^2}\right)\Biggl\{
 \cos^2\theta\, \Bigl[\frac{1}{c_W^2}\left(\frac{m_{h_1}^2}{m_{h_1}^2-M_Z^2}\right)\, \ln\, \frac{m_{h_1}^2}{M_Z^2}\\
&&-\left(\frac{m_{h_1}^2}{m_{h_1}^2-M_W^2}\right)\, \ln\, \frac{m_{h_1}^2}{M_W^2}\Bigr]+ \sin^2\theta\, \Bigl[\frac{1}{c_W^2}\left(\frac{m_{h_2}^2}{m_{h_2}^2-M_Z^2}\right)\nonumber \\
&&\times \ln\, \frac{m_{h_2}^2}{M_Z^2} - \left(\frac{m_{h_2}^2}{m_{h_2}^2-M_W^2}\right)\, \ln\, \frac{m_{h_2}^2}{M_W^2}\Bigr]\Biggr\} \ ,
\end{eqnarray}
following the feynman diagram method in Ref.~\cite{Barger:2007im}.
The SM T parameter $T^{\rm SM}$ can be recovered when $\cos\theta=1$. We plot the predictions of T parameter of the model in Fig.~\ref{fig:Tpar}. 
The quantity $\Delta T=T-T^{\rm SM}$ is subjected to the bound coming from the current global EW fit \cite{Baak:2014ora}:
$\Delta T=0.09\pm 0.13$. One can obtain severly constraints on the $\theta$ stronger with
increasing of $m_{h_2}$.

When the U(1) reduce to $Z_2$ with the pseudoscalar and U(1) breaking term being absent, the model reduce to the usual Higgs-portal real singlet model, and when the $Z_2$ is unbroken at zero temperature the $s$ field can serve as the dark matter candidate which can help the realization of singlet inflation~\cite{Lerner:2009xg}. Which will also be studied together with EWPT in this work.


\section{Ingredients of inflation, phase transition, and dark matter }
\label{sec:ipd}
\subsection{The scalar portal Inflation dynamics}

For the U(1) dark matter breaking model, the action in the Jordan frame is
\begin{eqnarray}
	S_J&&=\int d^4x \sqrt{-g}\Big[-\frac{M_{\rm p}^2}{2} R - \xi_h (H^\dagger H) R-\xi_s (\mathcal{S}^\dagger \mathcal{S}) R	 \nonumber\\
	&&+ \partial_\mu H^\dagger \partial^\mu H + \partial_\mu \mathcal{S}^\dagger \partial^\mu \mathcal{S} - V(H,\mathcal{S})\Big],
	\label{action}
\end{eqnarray}
where $\ds{M_{\rm p}}$ is the reduced Planck mass, $R$ is the Ricci scalar, $\xi_{\rm h,\rm s}$  define the non-minimal coupling of the $h,s$-field.

The quantum corrected effective Jordan frame Higgs potential at large field values ($h(s)$) can be written as
\be
V(h(s)) = \frac{1}{4} \lambda_{h(s)} (\mu) h(s)^4 \, ,
\ee
along the two-field potential evaluated along the higgs or singlet axis,
where the scale can be defined to be $\ds{\mu \sim \mathcal{O}(h) \approx h}$ in order to suppress the quantum correction. And the quartic couplings $\lambda_{h(s)}$ at planck scale can be obtained using the renormalization group equations given in Appendix.\ref{ap:A}. We impose quantum corrections to the potential and calculate the quantum corrections in the Jordan frame before performing the conformal transformation as in Ref.\cite{Elizalde:1993ew,Elizalde:2014xva}.
After the conformal transformation,
\be
\label{Omega}
\tilde{g}_{\mu\nu} = \Omega^2 g_{\mu\nu}, \quad \Omega^2\equiv 1+\frac{\xi_s s^2}{M_{\rm P}^2} + \frac{\xi_{\rm h} h^2}{M_{\rm P}^2}.
\ee
and a field redefinition
\be
\label{h_chi}
\frac{d\chi_{\rm h}}{dh} = \sqrt{\frac{\Omega^2+6\xi^2_{\rm h}h^2/M_{\rm P}^2}{\Omega^4}}, \quad \frac{d\chi_{\rm s}}{ds} = \sqrt{\frac{\Omega^2+6\xi^2_{\rm s}s^2/M_{\rm P}^2}{\Omega^4}},
\ee
we obtain
\be
\begin{aligned}
S_E =& \int d^4x \sqrt{-\tilde{g}}\bigg(-\frac{1}{2}M_{\rm P}^2R + \frac{1}{2}{\partial}_{\mu}\chi_{\rm h}{\partial}^{\mu}\chi_{\rm h} + \frac{1}{2}{\partial}_{\mu}\chi_{\rm s}
{\partial}^{\mu}\chi_{\rm s} \\
&+ f(\chi_{\rm s}, \chi_{\rm h}){\partial}_{\mu}\chi_{\rm h}{\partial}^{\mu}\chi_{\rm s} - U(\chi_{\rm s},\chi_{\rm h})  \bigg),
\label{EframeS}
\end{aligned}
\ee
where $U(\chi_{\rm s},\chi_{\rm h}) = \Omega^{-4}V(s(\chi_{\rm s}),h(\chi_{\rm h}))$ and
\be
f(\chi_{\rm s}, \chi_{\rm h}) = \frac{6\xi_{\rm h}\xi_{\rm s}}{M_{\rm P}^2\Omega^4}\frac{ds}{d\chi_{\rm s}}\frac{dh}{d\chi_{\rm h}}hs.
\ee

Basically, we can obtain $h-$ and $s-$inflations depends on if $\lambda_{\rm h}/\xi^2_{\rm h} \ll \lambda_{\rm s}/\xi^2_{\rm s}$ or $\lambda_{\rm h}/\xi^2_{\rm h} \gg \lambda_{\rm s}/\xi^2_{\rm s}$, see Ref.~\cite{Lerner:2009xg,Aravind:2015xst}. Then the kinetic terms of the scalar fields are canonical. We get to the Einstein frame by locally rescaling the metric by a factor $\ds{\Omega^2 = 1 + (\xi_h h^2 + \xi_s s^2) / M_{\rm pl}^2 \approx 1 + \xi_h h^2(s^2) / M_{\rm pl}^2 }$ with  $s(h)\sim 0$.
The non-canonical kinetic term for $h$ can be resolved by rewriting the inflationary action in terms of the canonically normalized field $\chi$ as
\eq{
S_{\rm inf} = \int d^4 x \sqrt{\tilde{g}} \left[ \frac{M_{\rm p}^2}{2}R + \frac{1}{2} \left( \partial \chi \right)^2 - U(\chi) \right] \;,
}
with the potential in terms of the canonically normalized field $\chi$ as
\eq{
U(\chi) = \frac{ \lambda_h  \left( h(\chi) \right)^4}{4 \Omega^4}  \, ~~{\rm or} ~~\, U(\chi) = \frac{ \lambda_s  \left( s(\chi) \right)^4}{4 \Omega^4}\;,
}
where the new field $\chi$ are defined by
\eq{
\frac{d \chi}{dh} \approx \sqrt{(1 + \xi_{h} h^2/M_{\rm p}^2 + 6 \xi_{h}^2 h^2 / M_{\rm p}^2)/( 1 + \xi_{h} h^2/M_{\rm p}^2 )^2  }\, ,
}
or
\eq{
\frac{d \chi}{ds}  \approx \sqrt{ (1 + \xi_s s^2/M_{\rm pl}^2 + 6 \xi_s^2 s^2 / M_{\rm pl}^2)/(1 + \xi_s s^2/M_{\rm pl}^2)^2 }\, .
}
for $h-$ or $s-$ inflations~\cite{Lerner:2009xg}.
Note that $\lambda_{h,s}$ and $\xi_{h,s}$ have a scale ($h(s)$) dependence.

 The slow-roll parameters are then given by
 \bea
\epsilon(\chi) = \frac{M_{\rm p}^2}{2} \left(\frac{dU/d\chi}{U(\chi)} \right)^2 \, , \qquad \eta(\chi) = M_{\rm p}^2 \left( \frac{d^2U/d\chi^2}{U(\chi)} \right)  \, .
\eea
The field value at the end of inflation $\chi_{\rm end}$ is obtained when $\ds{\epsilon= 1}$, and the horizon exit value $\chi_{\rm in}$ can be calculated with a fixed e-folding number between the two periods,
\be
N_{\rm e} = \int_{\chi_{\rm end}}^{\chi_{\rm in}} d\chi \frac{1}{M_{\rm p} \sqrt{2 \epsilon}}  \, .
\ee
Therefore, one can calculate the inflationary observables $n_s$ and $r$ with the slow-roll parameters at the $\chi_{\rm in}$ ,
\bea\label{eq:delta}
n_s  = 1 + 2 \eta - 6 \epsilon \, ,~
r = 16 \epsilon \, ,
\eea
Meanwhile, the amplitude of scalar fluctuations $\ds{\Delta_\mathcal{R}^2}$ can be calculated as
\eq{
\Delta_\mathcal{R}^2 = \frac{1}{24 \pi^2 M_{\rm p}^4}\frac{U(\chi)}{\epsilon} = 2.2\times10^{-9} \, .
}
The constraint on which is coming from CMB observations \cite{Ade:2015lrj}, which can be used to determine $\xi_{h,s}$.
We use the Plank bounds\cite{Ade:2015lrj} to require the values of $n_s$ and r to be $n_s = 0.9677 \pm0.0060$ at $1\sigma$ level and $r < 0.11$ at 95\% confidence level by assuming
$N_{\rm e}=60$. The slow-roll parameters $r$ are typically of order $\sim\mathcal{O}(10^{-2})$ for our cases.
The stability of the inflationary potential has been required by requiring conditions shown by Eq.~\ref{stab} from electroweak scale to planck scale using renormalization group equations (RGEs) listed in Appendix.~\ref{ap:B}.
The perturbativity of quartic coupling in the potential are also required following Ref.~\cite{Aravind:2015xst}.

 The non-minimal coupling for single field inflation are generally of order $\mathcal{O}(10^4)$, which might leads to possible unitarity problems at high scale around $\sim\mathcal{O}(10^{13})$ GeV~\cite{Burgess:2009ea,Barbon:2009ya,Barvinsky:2009ii,Bezrukov:2010jz,Burgess:2010zq,Hertzberg:2010dc}. While Ref.\cite{Calmet:2013hia,Burgess:2014lza,Bezrukov:2014ipa,Fumagalli:2016lls,Enckell:2016xse} argued that the SM Higgs inflation do not necessarily involve the problem.
The studies of Ref.\cite{Calmet:2013hia,Kahlhoefer:2015jma} indicates that the perturbative unitarity breaking can be healed by the additional singlet.
In this work, we do not address the issue.

\subsection{The cosmological phase transition calculation approach}

With the temperature cooling down, the universe can evolve from symmetric phase to the symmetry broken phase. The 
behavior can be studied with the finite temperature effective potential with particle physics models~\cite{Dolan:1973qd}.
Through which one can obtain the critical classics field value and temperature being $v_C$ and $T_C$ to characterize the critical phases. Roughly speaking, a SFOEWPT can be obtained when $v_C/T_C>1$, then the electroweak sphaleron process is
quenched inside the bubble and therefore one can obtain the net number of baryon over anti-baryon in the framework of EWBG. For the uncertainty of 
the value and possible gauge dependent issues we refer to Ref.~\cite{Patel:2011th}.

 The effective potential include the tree level Higgs potential described by Eq.~\ref{tree2}, the Coleman-Weinberg potential, and the finite temperature corrections take the form of~\cite{Arnold:1992rz},  
\begin{eqnarray}\label{eq:VfulT}
V(h,s,A,T)&=&V_0(h,s,A)+V_{CW} (h ,s ,A)+ V_1(h,s,A, T)\nonumber\\
&&+V_{\rm daisy}\left(h,s,A,T\right) \;,
\end{eqnarray}

With the field dependent masses being given in Appendix.~\ref{ap:A}, the one-loop Coleman-Weinberg scalar potential in $\overline{MS}$ and Landau gauge is,
\begin{eqnarray}
V_{CW} (h ,s ,A)&=& \frac{1}{{4 (4 \pi )^2}} M_i^4(h,s,A) \left(\log\frac{M_i^2(h,s,A)}{Q ^2} - c_i\right)\;,\nonumber\\
\end{eqnarray}
with $M_i$ to identify eigenvalues of scalar matrix, and other field dependent masses, here $n_{h_1,h_2,h_3,G^\pm,G^0,W^\pm,Z,t}=1,1,1,2,1,6,3,-12$, and $c_{W^\pm,Z}=5/6$ with others $c_{i}$ being $3/2$. The running scale $Q$ is chosen to be $Q=246.22$ GeV in the numerical analysis process. The finite temperature corrections to the effective potential at one-loop are given by~\cite{Dolan:1973qd},
\begin{eqnarray}
 V_1(h_,s,A, T) = \frac{T^4}{2\pi^2}\, \sum_i n_i J_{B,F}\left( \frac{ M_i^2(h,s,A)}{T^2}\right),
\end{eqnarray}
where the functions $J_{B,F}(y)$ are
\begin{eqnarray}
 J_{B,F}(y) = \pm \int_0^\infty\, dx\, x^2\, \ln\left[1\mp {\rm exp}\left(-\sqrt{x^2+y}\right)\right],
\end{eqnarray}
with the upper (lower) sign corresponds to bosonic (fermionic) contributions. Here, the above integral $J_{B,F}$ can be expressed as a sum of them second kind modified Bessel functions $K_{2} (x)$~\cite{Anderson:1991zb},
\begin{eqnarray}
J_{B,F}(y) = \lim_{N \to +\infty} \mp \sum_{l=1}^{N} {(\pm1)^{l}  y \over l^{2}} K_{2} (\sqrt{y} l)\;.
\end{eqnarray}
Last but not least, the resummation of \textit{ring }(or\textit{\ daisy}) diagrams are also crucial for the evaluation of $v_C$ and $T_C$ with the finite temperature effective potential~\cite{Carrington:1991hz},
\begin{eqnarray}
V_{\rm daisy}\left(h,s,A,T\right) =\frac{T}{12\pi }\sum_{i} n_{i}\left[\left( M_{i}^{2}\left( h,s,A\right) \right)^{\frac{3}{2}} -\left( M_{i}^{2}\left(h,s,A,T\right) \right)^{\frac{3}{2}}\right]\;,\nonumber\\
\end{eqnarray}
where $M_{i}^{2}\left( h,s,A,T\right) $ are the eigenvalues of the full bosonic mass matrix with thermal corrected 
effects being taken into accounted($M_{i}^{2}\left( h,s,A,T\right) =M_{i}^{2}\left( h,s,A\right)+M_x^2(T)$), the thermal correction masses $M_x^2(T)$ are given in Appendix.~\ref{ap:A}.

Then the critical parameters of EWPT can be calculated when there are two degenerate vacuums with a potential barrier. Due to rich vacuum structures of the potential at finite temperature, there can be one-step or multi-step phase transitions. A SFOEWPT can be realized in the first or the second step in the two-step scenario. When the U(1) reduces to $Z_2$ symmetry with the 
$A$ absent, one return to the usual  
real singlet case. 
As studied previously in Ref.~\cite{Curtin:2014jma}, the one-step mode calls for a larger quartic coupling between $h$ and $s$,
two-step phase transition can happens at a relatively smaller quartic coupling between $h$ and $s$. When the singlet $s$ serve as the 
dark matter candidate, the two-step mode indicates the $Z_2$ symmetry is broken at some higher finite temperatures and restored at some lower and zero temperatures.
For the complex singlet case, the studies of Ref.~\cite{Jiang:2015cwa,Chiang:2017nmu} shows that the dark matter masses effects on the evolution of effective potential with temperature cooling down is negligible. And for simplicity, we do not expect the $A$ field get vacuum exception value (VEV) at finite temperature, and focus on the case where the vacua can happen along the $h$ and/or $s$ directions with the temperature decreasing. Again, to obtain a strong first order phase transition together with the successful inflation, we focus on 
the two-step phase transition here since the inflation requires small quartic scalar couplings as studied previously in Ref.~\cite{Lerner:2009xg,Aravind:2015xst} and this work.
  
\begin{figure}[!htp]
\begin{center}
\includegraphics[width=0.23\textwidth]{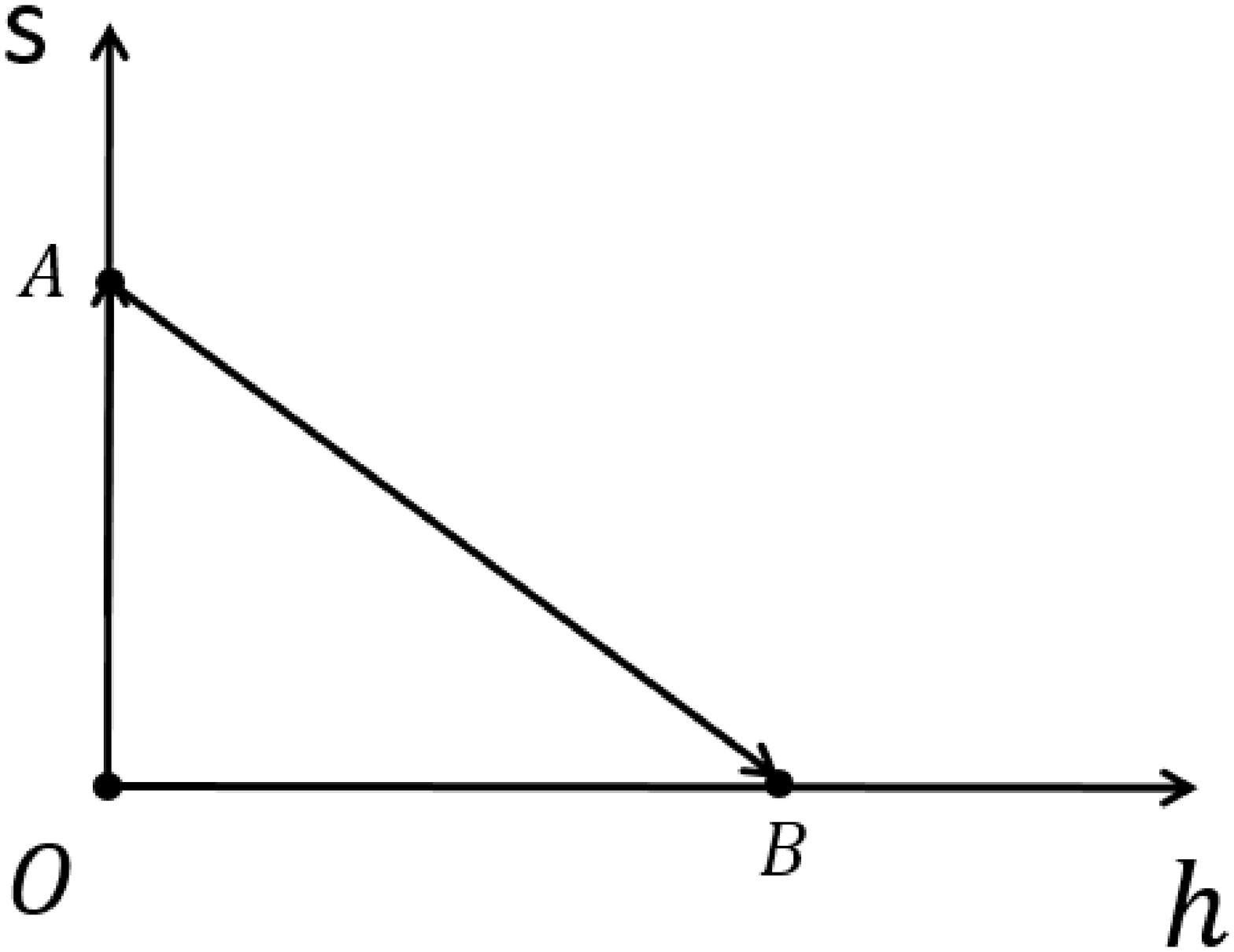}
\includegraphics[width=0.23\textwidth]{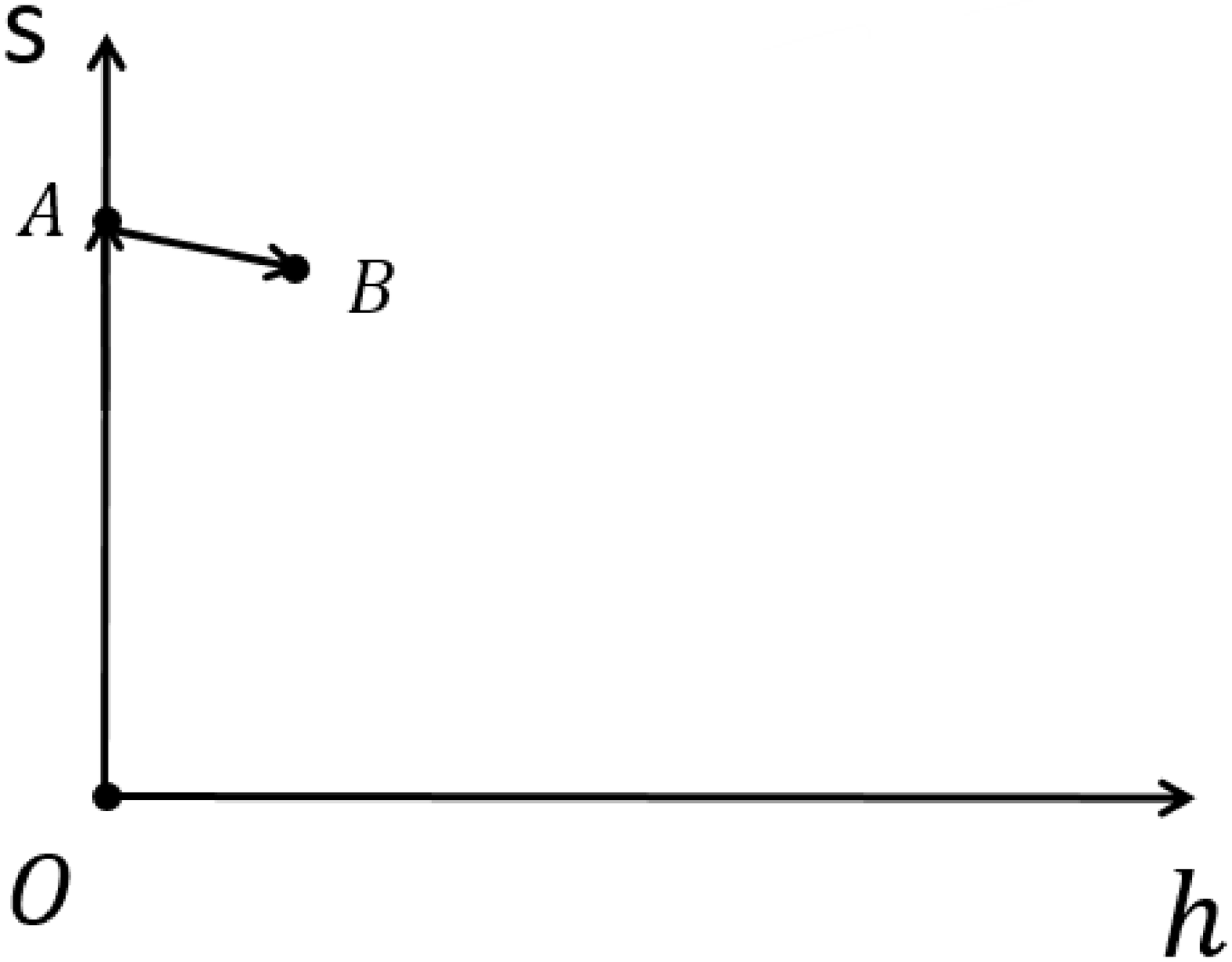}
\caption{EWPT types in Higgs-portal DM and U(1) breaking DM models for left and right panels. }
\label{fig:EWPTtype}
\end{center}
\end{figure}

The phase transition types of the complex and real singlet cases are shown by Fig.~\ref{fig:EWPTtype}.
The left panel is for the real singlet case.
In this case, the form of the finite temperature effective potential
reduce from Eq.~\ref{eq:VfulT} to $V(h,s,T)$ without $A$ contributions.
At the critical temperatures for different set of quartic couplings and dark matter masses, one have two local minimums, wherein
\begin{eqnarray}
&&V(0,s_C^A,T_C)=V(v_C^B,0,T_C)\;,\nonumber\\
&&\frac{dV(0,s,T_C)}{ds}|_{s=s_{C}^A}=0,~\frac{dV(h,0,T_C)}{dh}|_{h=v_{C}^B}=0\;.
\end{eqnarray} 
The right panel of Fig.~\ref{fig:EWPTtype} is for the phase transition type of the complex singlet case.
Considering the mixing of $h$ and $s$ at zero temperature as shown in the Sec.~\ref{sec:mod}, one can replace the 
classical fields $h,s$ by $h_{1,2}$ accompanied by the mixing angle $\theta$. 
The critical temperature and critical field value can be evaluated through,
\begin{eqnarray}
&&V(0,s_C^A,\theta,0,T_C)=V(v_C^B,s_C^B,\theta,0,T_C)\;,\nonumber\\
&&\frac{dV(h,s,\theta,0,T_C)}{ds}|_{h=v_{C}^B,s=s_{C}^B}=0\;,\nonumber\\
&&\frac{dV(h,s,\theta,0,T_C)}{dh}|_{h=v_{C}^B,s=s_{C}^B}=0\;,\nonumber\\
&&\frac{dV(0,s,\theta,0,T_C)}{ds}|_{s=s_{C}^A}=0\;,
\end{eqnarray} 
when two degenerate vacuum with a potential barrier structures shows up for a set of $m_{h_1,h_2,A}$ and $\theta$ at finite temperatures. Here, we assume the classical field A do not get VEV at any temperature, as aforementioned. 

\subsection{The Higgs-portal dark matter candidate}
\subsubsection{Dark matter relic density}

\begin{figure}[!ht]
\includegraphics[width=0.15 \textwidth]{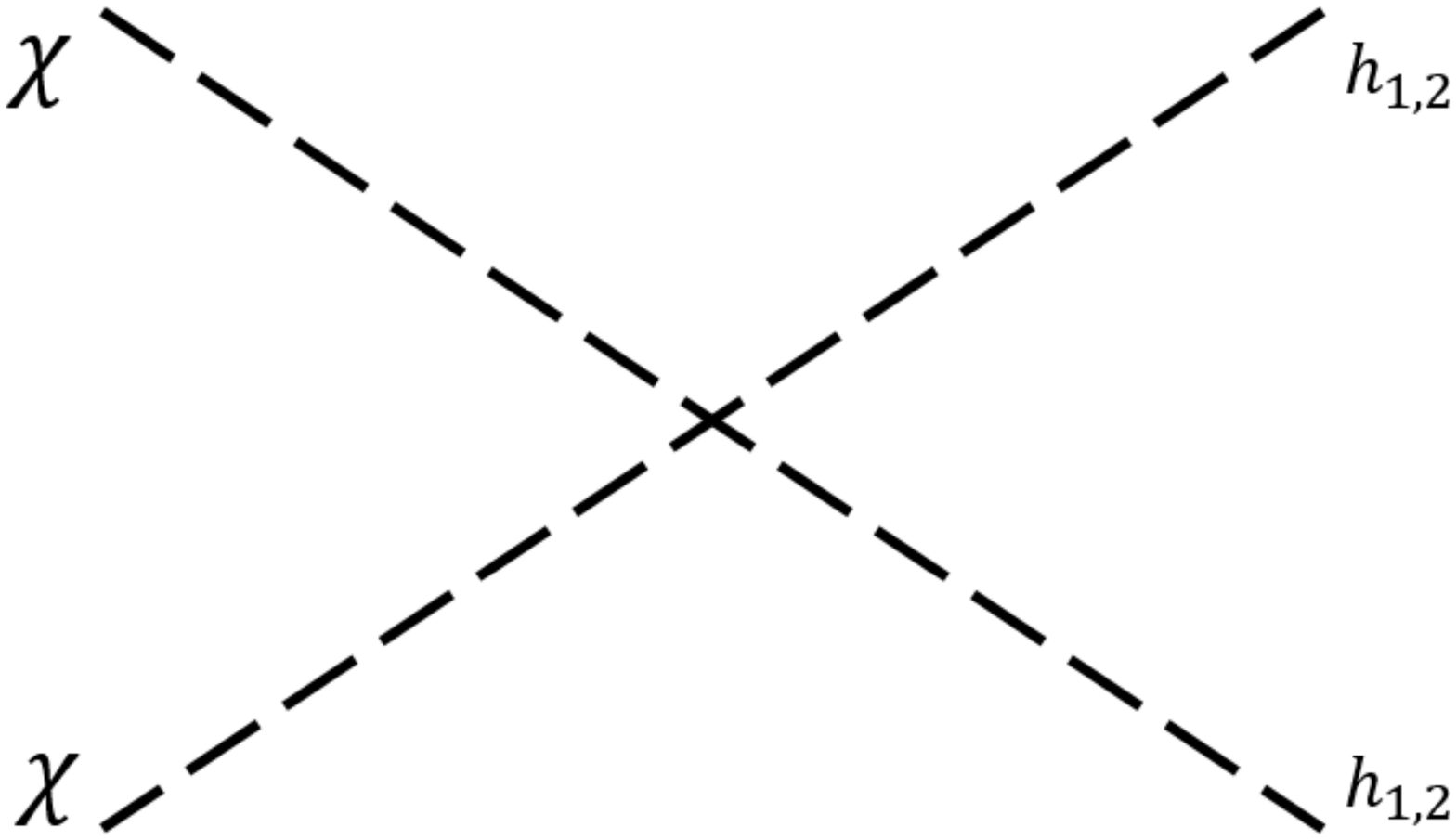}
\includegraphics[width=0.15 \textwidth]{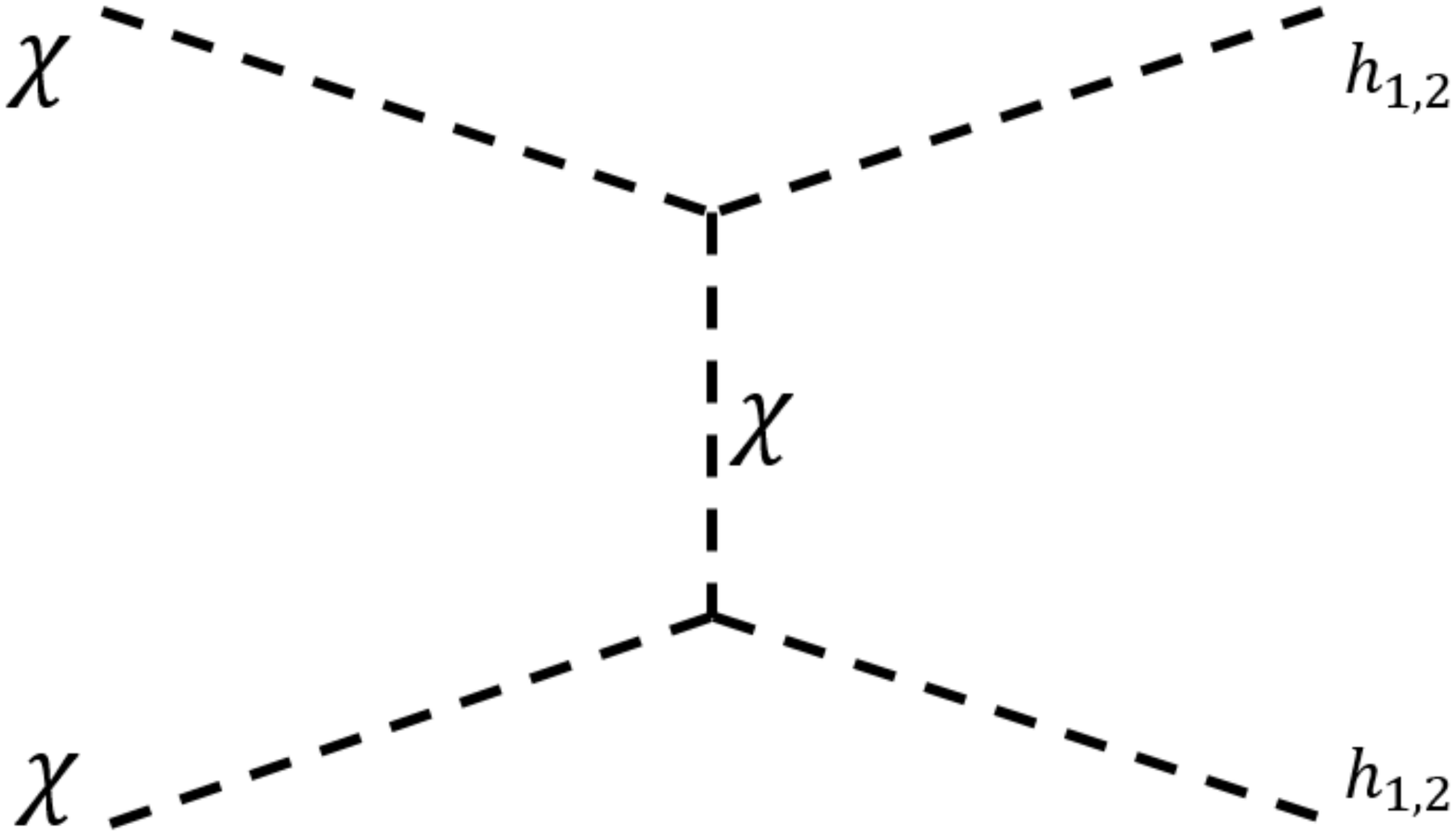}
\includegraphics[width=0.15 \textwidth]{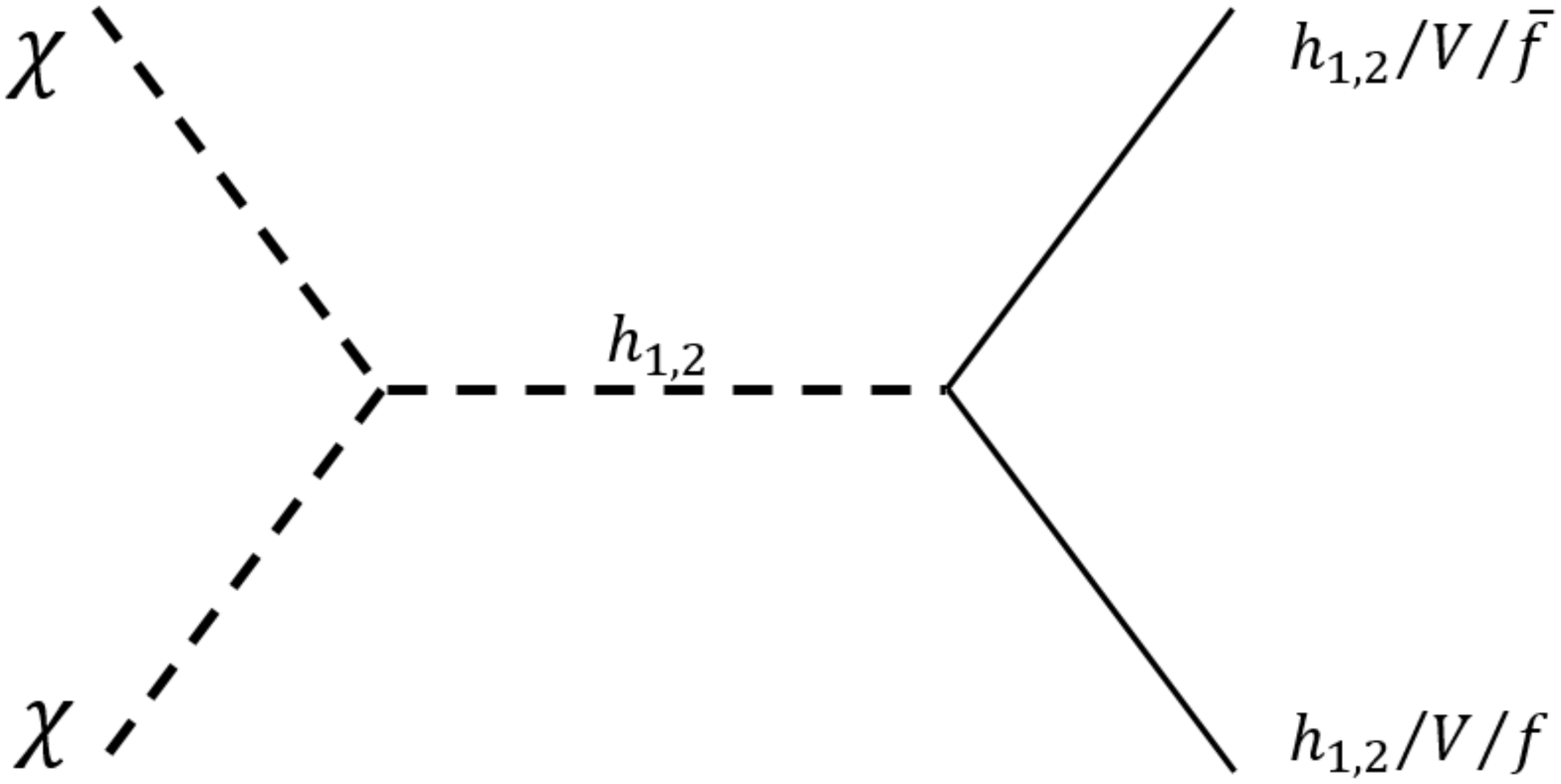}
\caption{Annihilation channels of the U(1) breaking dark matter model.}\label{fig:ann}
\end{figure}

The dark matter annihilation process are shown in Fig.~\ref{fig:ann}. The thermal averaged dark matter pairs annihilation cross sections are given by,
\begin{eqnarray}
\sigma v_{r} = \langle\sigma v\rangle_{ZZ} + \langle\sigma v\rangle_{WW }+ \langle\sigma v\rangle_{f\bar{f}}+\langle \sigma v\rangle_{h_ih_j},
\end{eqnarray}
with $i,j=1,2$ and these contributions are listed in Appendix.\ref{ap:C}, these formula simply reduced to the formula of Ref.\cite{Lerner:2009xg} for the gauge singlet, that gives rise to the usual Higgs-portal real singlet scalar dark matter scenario. With the thermal averaged dark matter annihilation cross sections at hand, we calculate the relic density with the method of~\cite{Kolb}, which is checked to be around percent level descrepency with the micrOMEGAs~\cite{micromegas}. The current value of the relic abundance of dark matter $\Omega_{dm} h^2 \approx 0.12$~\cite{Ade:2015xua}. 

\subsubsection{Dark matter direct detection}

Due to the mixing of h and s, one can expect the $h_{1,2}$ mediated diagrams contribute to the spin independent cross section~\cite{Barger:2010yn,Gonderinger:2012rd},
\begin{eqnarray}
\sigma_{SI}&=&\frac{m_{p}^{4}}{2\pi v^{2}(m_{p}+M_{A})^{2}}\left(\frac{g_{h_1AA}g_{h_{1}}}{m_{h_{1}}^{2}}+\frac{g_{h_{2}AA}g_{h_{2}}}{m_{h_{2}}^{2}}\right)^{2}\nonumber\;\\
&&\times
\left(f_{pu}+f_{pd}+f_{ps}+\frac{2}{27}(3f_{G})\right)^{2},
\label{eq:sip}
\end{eqnarray}
where $m_{p}$ is the proton mass and $v=246$ GeV is the SM-Higgs VEV, and the central values of strengths of the hadronic matrix elements, $f$, are given by~\cite{Ellis:2000ds}
\begin{equation}
f_{pu} = 0.02,\quad f_{pd} = 0.026,\quad f_{ps} = 0.118,\quad f_{G}=0.836.\nonumber\;
\end{equation}
Here, one note that cancellation between the two parts can happen depends on mixing angle and the heavy Higgs masses. The behavior
would relax the parameter spaces being bounded by the direct detection experiments, especially after the strongest bounds coming from 
the XENON1T~\cite{Aprile:2017iyp}. The direct detection bounds are set by rescale the dark matter relic abundance when the dark matter is undersaturated~\cite{Gonderinger:2012rd}. It should be note that after take into account the over-estimated Higgs-nucleon interaction coupling $f_N$~\cite{Hoferichter:2017olk}, the parameter spaces allowed by direct detection experiments can be extended.

\section{Results}\label{sec:res}

Before the numerically analysis of the inflation and EWPT with the assistance of the complex scalar sector, we first comment on the temperatures of reheating, electroweak phase transition, and dark matter freeze out briefly.
The parametric resonance of the oscillating Higgs field to W bosons(singlet scalar) via $|H|^2|W|^2$($|H|^2 |S|^2$) can help the the Higgs (singlet) inflation reheating occurs\cite{Bezrukov:2008ut,Lerner:2009xg}.
Ref.~\cite{Tenkanen:2016idg} studied typical small $\lambda_{s}\sim\mathcal{O}(10^{-9}-10^{-2})$ for singlet inflation reheating. 
Ref.~\cite{Lerner:2011ge} studied $s-$inflation reheating with large $\lambda_s$, which we can apply to our analysis. 
The reheating can happen through the stochastic resonance to the Higgs bosons or the production of the s-inflaton excitations
in the case of the complex singlet scalar model. While, for the higgs-portal real singlet dark matter model, the reheating can occurs due to the production of the $s-$inflaton excitations where $\lambda_s>0.019$. 

After the reheating, the universe can undergo cosmological phase transition at around the temperature of $T_C\sim\mathcal{O}(10^2)$ GeV, then the symmetry breaking happens after that, i.e., EWPT happens.
As noted by Ref~\cite{McDonald:1993ex}, the freeze out temperature($T_{fs}$) should be smaller than the EWPT temperatures due to the thermal averaged annihilation cross section(see Eq.\ref{eq:ann}) in our relic density evaluation process is T-independent. In particular, indeed in the the dark matter mass regions smaller than $\sim\mathcal{O}(10)$ TeV, one always can have the $T_{fs}\leq T_C$. Therefore, one can calculate the relic density self-consistently in that dark matter mass regions.
Furthermore, at this stage the non-minimal gravity couplings effect is negligible, for the typical case of dark matter decay through the non-minimal gravity couplings we refer to Ref~\cite{Cata:2016epa,Cata:2016dsg}.  

As studied previously by Ref.~\cite{Curtin:2014jma}, the SFOEWPT happens more easy through two-step for smaller quartic coupling $\lambda_{hs}$ in the real singlet case, as shown in the left panel of Fig.~\ref{fig:EWPTtype}. Where, the $Z_2$ symmetry can broken at finite temperature and restored at zero temperature, see also~\cite{Cline:2012hg,Cline:2017qpe}. For the complex singlet case, we find that only in the two-step scenario(see right panel of the Fig.~\ref{fig:EWPTtype}), the phase transition can be strong first order for a smaller quartic Higgs couplings(see Ref.~\cite{Jiang:2015cwa,Chiang:2017nmu}) in order to account for the inflation. On the other hand, a larger coupling of the SM Higgs and  the complex singlet can easily be excluded by the direct detection of dark matter. Instead, the U(1) breaking complex singlet case is more easy to situate the dark matter framework than the Higgs portal real singlet scalar case due to the exist of the cancellation can relax  the direct detection bounds on parameter spaces~\cite{Gross:2017dan}.

\begin{figure}[!ht]
\begin{center}
\includegraphics[width=0.4\textwidth]{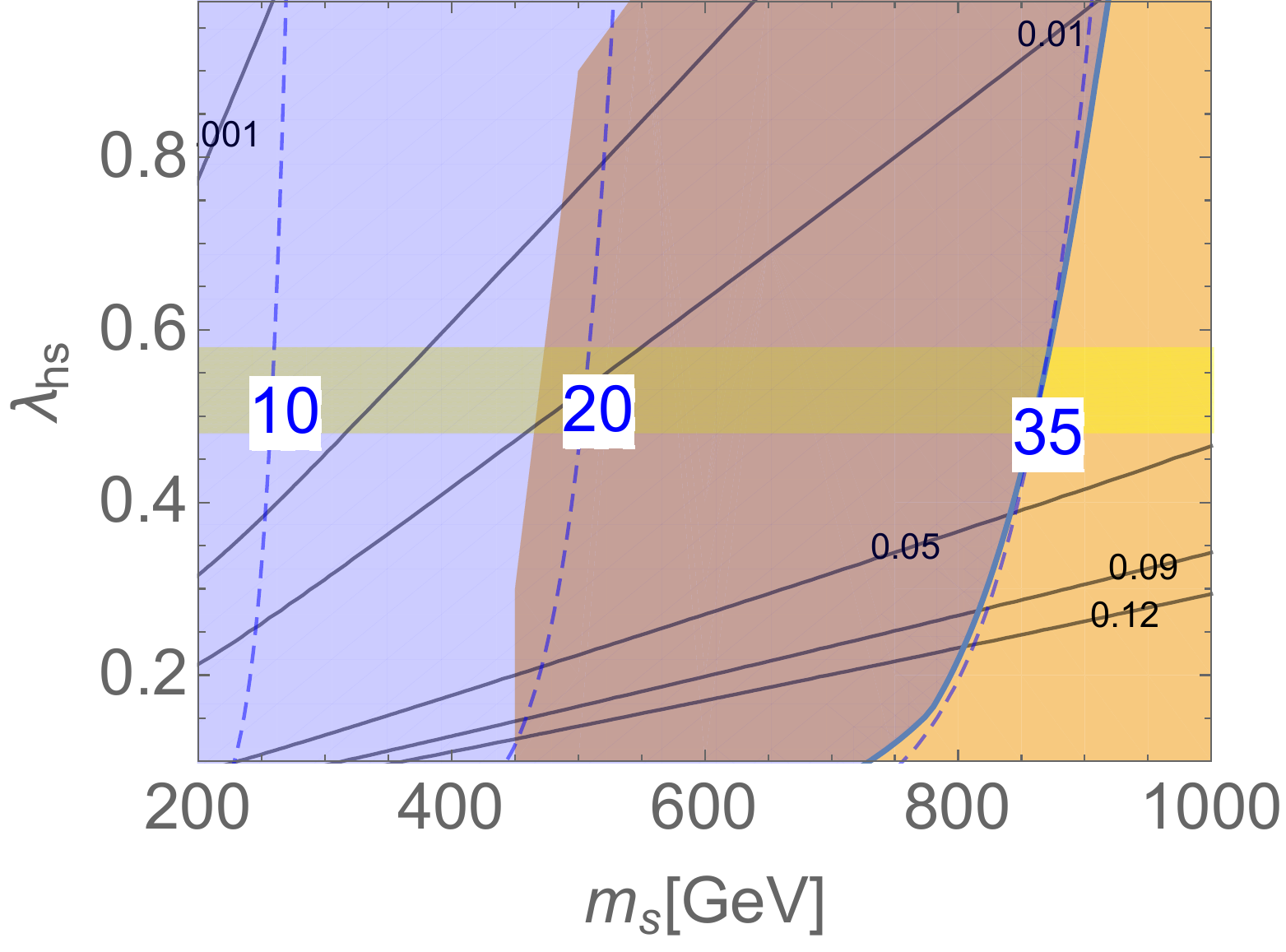}
\caption{Higgs-portal DM scenario. The freeze out temperature is shown in dashed line. The solid line is for the DM relic density, and the direct detection bounds from Xenon1T are marked by blue regions. In the yellow and orange regions we can obtain $h(s)$ inflation and strong first order EWPT.}
\label{fig:singletinflation}
\end{center}
\end{figure}

For real singlet as dark matter scenario, as shown in Fig.~\ref{fig:singletinflation}, in the parameter regions where SFOEWPT can be obtained the dark matter masses are larger than 450 GeV, the successful realization of inflation occurs around $\lambda_{hs}\sim 0.5$, larger(smaller) $\lambda_{hs}$ is excluded by the stability bounds together the requirement of perturbativity of quartic coupling.
The dark matter direct detection experiment Xenon 1T~\cite{Aprile:2017iyp} exclude the parameter spaces for dark matter mass $m_S$ being smaller than $\sim$700 GeV. The parameter spaces that can explain inflation, and dark matter relic density is around $m_S\sim\mathcal{O}(1)$TeV.

\begin{figure}[!ht]
\begin{center}
\includegraphics[width=0.23\textwidth]{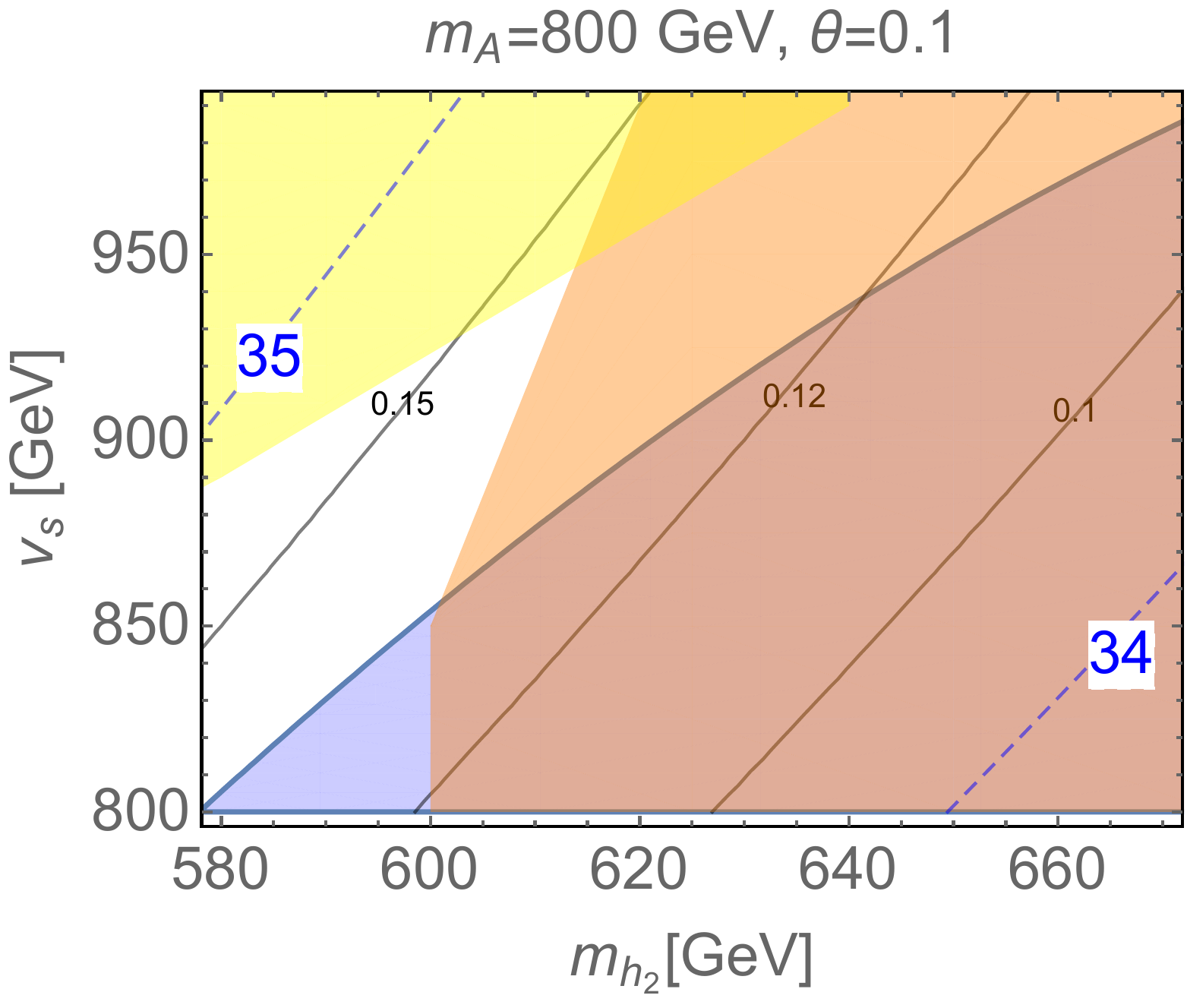}
\includegraphics[width=0.23\textwidth]{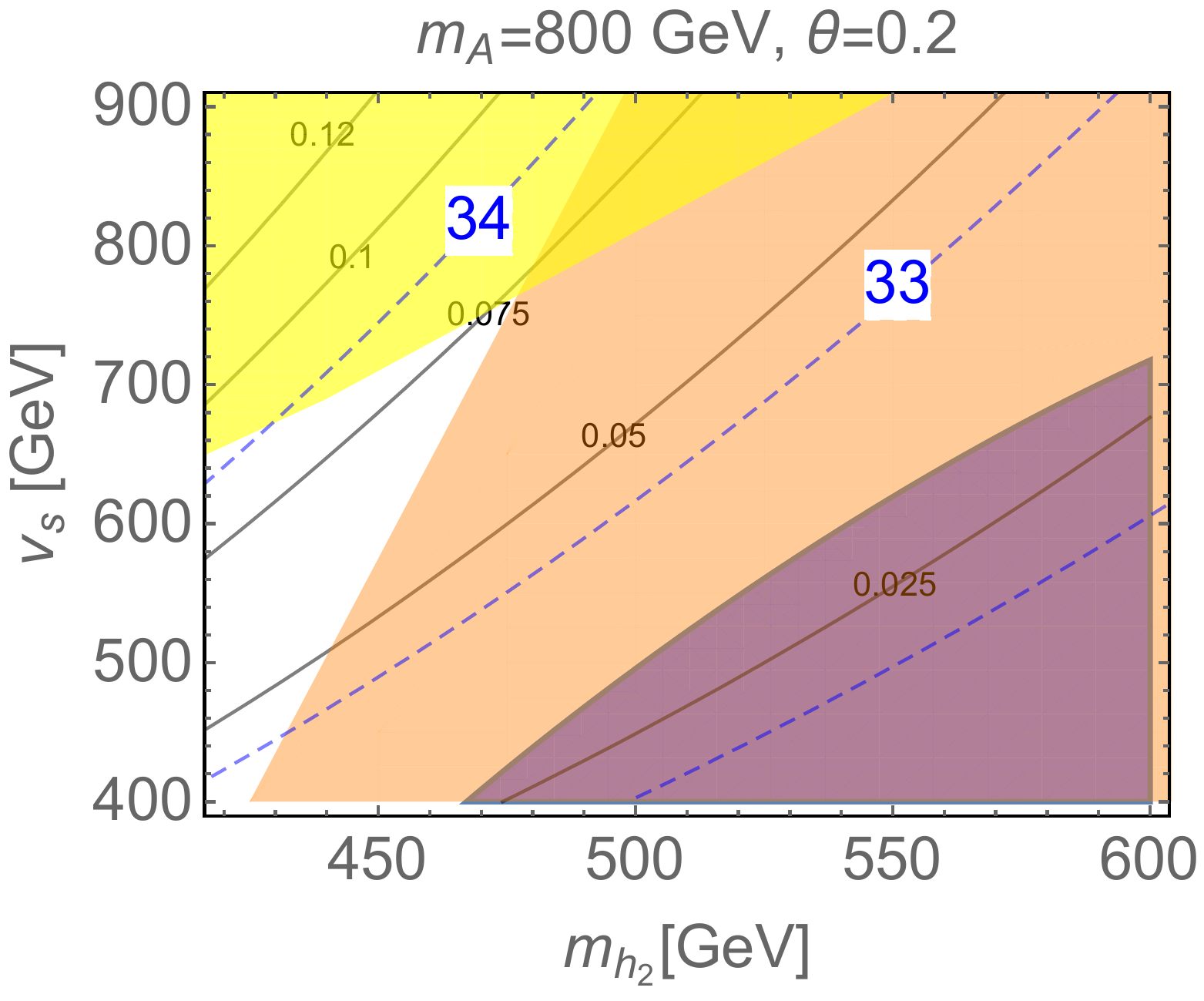}
\caption{ U(1) breaking pseudoscalar DM model for left and right panels. The color-codes are the same as Fig.~\ref{fig:singletinflation}.}
\label{fig:U1inflation}
\end{center}
\end{figure}

For the U(1) breaking complex singlet model, where the dark matter candidate is the pseudoscalar. As in the usual Higgs-portal DM case,
a larger dark matter masses leads to a larger suppression of annihilation cross section and the DM-nucleon scattering cross section, and therefore results in correct magnitude of relic density that escaping the direct detection
bounds. Here, the dark matter mass is totally determined by the U(1) breaking parameter, i.e., $\mu_b$. It should be note that which is independent of the inflation dynamics since it do not enter into RGEs of quartic couplings. The effects of dark matter masses are founded to be negligible for the two-step EWPT process within this framework, as explored in Ref.~\cite{Jiang:2015cwa,Chiang:2017nmu}.
We present the results with $m_A=800$ GeV and $\theta=0.1,0.2$ in Fig.~\ref{fig:U1inflation}. Where a larger $v_s$(leads to ) and smaller $m_{h_2}$ is required to successfully implement inflation. This is mostly because that
a larger $\theta$ is accompanied by a smaller $v_s$ and larger $m_{h_2}$ in order do not violate the perturbativity of quartic coupling and can make the vacuum stable up to the Planck scale, as can be figured out
through the comparison of the parameter region of the two panels. A SFOEWPT for a larger $\theta$ can be obtained with a smaller $m_{h_2}$ and $v_s$. The critical temperature of the SFOEWPT is found to be $\mathcal{O}(10^2)$ GeV, which is larger than the freeze out temperatures. The magnitude of the relic density can increase from under-saturate to oversaturate with increasing (decreasing )of $v_s$ ($m_{h_2}$) since then one have a smaller annihilation cross section of dark matter pairs. In the left panel of Fig.~\ref{fig:U1inflation}, with a mixing angle of $\theta=0.1$, the dark matter relic density is oversaturated though both the successful inflation and a SFOEWPT can be accomplished. In the right panel, we show the mixing angle increase to be $\theta=0.2$, it illustrates that the dark matter relic density is undersaturated in the parameter region where the successful inflation and a SFOEWPT can occur. The two panels indicates the correct relic density together with the successful inflation and a SFOEWPT can be obtained when the mixing angle $0.1<\theta<0.2$.
With the increasing of $\theta$, the smaller $v_s$ regions would be excluded by the Xenon 1T experiment since a larger $\theta$ can lead to a lower magnitude of cancellation for the direct detection in the parameter spaces of the same $v_s$ or $m_{h_2}$, as indicted by Eq.~\ref{eq:sip}.

\section{Conclusions and discussions}\label{sec:condis}

In this work, we studied the inflation dynamics and cosmological phase transition in the framework of the usual Higgs-portal singlet scalar model and the complex singlet scalar extended SM with the broken global U(1).
For the Higgs-portal singlet scalar model, the dark matter relic abundant is found to be undersaturated within the parameter spaces where successful inflation and a SFOEWPT can occur.
In the complex singlet scalar extended model, the simultaneously address of successful inflation and strong first order electroweak phase transition can be obtained with the dark matter masses being around $800$ GeV when the mixing angle $0.1<\theta<0.2$. In comparison with the usual Higgs-portal singlet scalar dark matter case, the cancellation of direct detection in the scenario of complex singlet scalar can be more easy to address the two problems depending on the mixing angle and the heavy Higgs masses. 

With an additional CP-violation phase by implementing with CPV dim-6 operator in the complex singlet model as in Ref.~\cite{Jiang:2015cwa}, the flatness of the potential will not be destoryed and the baryon asymmetry can be generated during the EWPT process within the framework of EWBG mechanism. And then, inflation, EWBG, as well as dark matter relic density all can be addressed despite a larger dark matter masses are needed in comparison with Ref.~\cite{Jiang:2015cwa,Beniwal:2017eik} in order for relatively small quartic scalar couplings required by the successful inflation. 
 A recent bubble nucleation analysis in the gauge singlet scalar case(respect the $Z_2$ symmetry at zero temperature) performed by Ref.~\cite{Kurup:2017dzf} indicated that the two-step phase transition parameter spaces there is strongly restricted. That make detection of the bubble nucleation (proceeded by strong first order phase transition) generated gravititional wave signal with typical frequency of $\mathcal{O}(10^{-4}-10^{-2})$Hz by eLISA more harder. The study of more general gauge singlet scalar case with no discrete symmetry shows that the GW signal of spectrum frequency of $\mathcal{O}(10-100)$Hz generated by
 a high scale bubble nucleation with temperature $\sim 10^7$ GeV can be probed by aLIGO. 
 The search of the parameter spaces in Fig.~\ref{fig:U1inflation} can be performed through the resonant heavy Higgs search at LHC and SPPC~\cite{Arkani-Hamed:2015vfh,Kotwal:2016tex,Huang:2017jws}. 
 We left the GW signals and the collider searches of the parameter spaces wherein inflation, SFOEWPT, and dark matter can be addressed simultaneously to future studies.

\section*{ACKNOWLEDGMENTS}
LGB thank Hyun Min Lee for helpful discussions on inflations.
The work of LGB is partially supported by the National Natural Science Foundation of China (under Grant No. 11605016), Basic Science Research Program through the National Research Foundation of Korea (NRF) funded by the Ministry of Education, Science and Technology (NRF-2016R1A2B4008759), and Korea Research Fellowship Program through the National Research Foundation of Korea (NRF) funded by the Ministry of Science and ICT (2017H1D3A1A01014046).

\section{Appendix}

\subsection{Field dependent masses and thermal masses}\label{ap:A}
The field dependent masses are given by,
\begin{eqnarray}
\label{eq:Mhschi}
M(h,s,A)  = \left( {\begin{array}{*{20}{c}}
m_{hh}&m_{hs}&m_{hA}\\
m_{hs}&m_{ss}&m_{sA}\\
m_{hA}&m_{sA}&m_{AA}
\end{array}} \right).
\end{eqnarray}
with
\begin{eqnarray}
m_{hh}&=&{\frac{1}{2}(6\lambda_h{h^2} -2 \mu_h^2 + \lambda_{hs}({s^2} + {A ^2}))},\\
m_{hs}&=&{\lambda_{hs}{hs}},\\
m_{hA}&=&{\lambda_{hs}hA },\\
m_{ss}&=&{\frac{1}{2}\lambda_{hs}{h^2} -\mu_s^2-\mu_{b}^2 + \lambda_s(3 s^2 + {A ^2})},\\
m_{sA}&=&{2\lambda_s sA },\\
m_{AA}&=&{\frac{1}{2}\lambda_{hs}{h^2} - \mu_s^2+ \mu_b^2}+ \lambda_s({3A ^2}  + s^2).
\end{eqnarray}
The mass matrix Eq.\ref{eq:Mhschi} can be diagonalized, with eigenvalues values being $M^2_{1,2,3}$, other field dependent masses are,
\begin{eqnarray}
M^2_{G^0}(h, s,A ) &=& \frac{1}{2}(2\lambda_h {h^2}+ \lambda_{hs}{A ^2} - 2\mu_h^2 + \lambda_{hs} s^2)\;,\\
M^2_{G^\pm}(h, s ,A )&=& \frac{1}{2}(2\lambda_h {h^2}+ \lambda_{hs}{A ^2} - 2\mu_h^2 + \lambda_{hs} s^2)\; ,\\
M^2_{t}(h )&=& \frac{{g_t^2{h^2}}}{2}\;,\\
M^2_{Z}(h ) &= &\frac{1}{4}(g_1^2 + g_2^2){h^2}\;,\\
M^2_{W}(h )&= &\frac{{g_2^2{h^2}}}{4}\;.
\end{eqnarray}
The thermal masses/corrections in the U(1) breaking model are given by,
\begin{eqnarray}
M_{h_i} ^2(T) &=& \frac{{g_1^2{T^2}}}{{16}} + \frac{{3g_2^2{T^2}}}{{16}} + \frac{{g_t^2{T^2}}}{4} + \frac{{\lambda_h{T^2}}}{2} + \frac{{\lambda_{hs}{T^2}}}{12},\\
M_{G^{0,\pm}}^2(T) &=& M^2_{h_i} (T),\\
M_s^2(T )&=& \frac{{\lambda_{hs}{T^2}}}{6} + \frac{{\lambda_s{T^2}}}{3},\\
M^2_A (T)& = &\frac{{\lambda_{hs}{T^2}}}{6} + \frac{{\lambda_s{T^2}}}{3}.
\end{eqnarray}
for the scalar fields, the gauge fields thermal masses can be found in Ref.~\cite{Cai:2017tmh}.

\subsection{ Beta Functions} \label{ap:B}
The one-loop beta functions for the various parameters can be found in the Higgs-portal real singlet case~\cite{Aravind:2015xst} except that
for the complex singlet scenarios, the scalar quartic coupling beta functions are replaced by
\begin{eqnarray}
\beta_ {\lambda_{h}}&=& \frac{3 g_1^4}{128 \pi ^2}+\frac{3g_{1}^2 g_{2}^2}{64 \pi ^2}-\frac{3 g_{1}^2\lambda_{h}}{16 \pi ^2}+\frac{9 g_{2}^4}{128 \pi ^2}-\frac{9g_{2}^2\lambda_h}{16 \pi ^2}-\frac{3 g_{t}^4}{8 \pi ^2}\nonumber\\
&&+\frac{3 g_{t}^2\lambda_{h}}{4 \pi ^2}+\frac{3 \lambda_{h}^2}{2 \pi ^2}+\frac{\lambda_{hs}^2}{16 \pi ^2}\; ,
\\
\beta_ {\lambda_s}&=& \frac{\lambda_{hs}^2}{8 \pi ^2}+\frac{5 \lambda_s^2}{4 \pi ^2}\; ,
\\
\beta_{ \lambda_{hs}}&=& -\frac{3 g_{1}^2\lambda_{hs}}{32 \pi ^2}-\frac{9 g_{2}^2 \lambda_{hs}}{32 \pi ^2}+\frac{3g_{t}^2\lambda_{hs}}{8 \pi ^2}+\frac{3\lambda_{h} \lambda_{hs}}{4 \pi ^2}+\frac{\lambda_{hs}^2}{4 \pi ^2}+\frac{\lambda_{hs}\lambda_{s}}{2 \pi ^2}\; .\nonumber\\
\end{eqnarray}
We use the electroweak scale values of the various couplings consistent with \cite{Buttazzo:2013uya} at the initial conditions of the RGEs.

\subsection{DM annihilations cross sections}
\label{ap:C}
The relevant cubic and quartic interaction couplings being given by,
\begin{eqnarray}
g_{h_1} &=& \cos\theta,~g_{h_2} =- \sin\theta,\nonumber\\
g_{h_1A A}  &=& \lambda_{hs}v\cos\theta+ \lambda_s v_s \sin\theta\,,\nonumber\\
g_{h_2A A}  &=& 2\lambda_{s}v_s\cos\theta - \lambda_{hs}v\sin\theta,\nonumber\\
g_{h_1h_1h_1} &=& 3[2\lambda_{h}v (\cos \theta)^3 + \lambda_{hs}v(\sin\theta)^2\cos\theta \nonumber\\
&&+ \lambda_{hs}v_s\sin\theta (\cos\theta)^2 +2 \lambda_{s}v_s(\sin \theta)^3],\nonumber\\
g_{h_2h_1h_1 }&=& 2(\lambda_{hs} - 3\lambda_h)v\sin\theta (\cos\theta)^2 - \lambda_{hs}v(\sin \theta)^3 \nonumber\\
&&+ 2(3\lambda_{s}- \lambda_{hs})v_s(\sin\theta)^2\cos\theta + \lambda_{hs}v_s(\cos\theta)^3,\nonumber\\
g_{h_2h_2h_2 }&=& 3( - 2\lambda_{h}v(\sin\theta)^3 - \lambda_{hs}v\sin\theta (\cos\theta)^2\nonumber\\
&&+ \lambda_{hs}v_s(\sin\theta)^2\cos\theta + \lambda_{s}v_s(\cos\theta)^3),\nonumber
\end{eqnarray}
\begin{eqnarray}
g_{h_1h_2h_2} &=& v(3\lambda_h - \lambda_{hs})(\sin 2\theta) \sin\theta + \lambda_{hs}v (\cos\theta)^3 \nonumber\\
&&+2 (3\lambda_{s} - \lambda_{hs})v_s\sin\theta (\cos\theta)^2 + \lambda_{hs}v_s(\sin\theta)^3,\nonumber\\ 
g_{ h_1h_2A A} &=& (2\lambda_{s} - \lambda_{hs})\sin \theta \cos \theta ,\nonumber\\
g_{ h_2h_2A A} &=& \lambda_{hs}(\sin \theta)^2 + 2\lambda_{s}(\cos\theta)^2,\nonumber\\
g_{ h_1h_1A A} &=& \lambda_{hs}(\cos \theta)^2 +2 \lambda_{s}(\sin\theta)^2.
\end{eqnarray}

With these couplings and the propogators of $h_{1,2}$,
\begin{eqnarray}
D_{h_1} &=& (4m_A^2 - m_{h_1}^2) +\rm{I} \Gamma _{h_1}m_{h_1}  \;,\\
D_{h_2} &=& (4m_A^2 - m_{h_2}^2) + \rm{I} \Gamma _{h_2}m_{h_2} \; ,
\end{eqnarray}
the thermal averaged annihilation cross sections are given by,
\begin{widetext}
\begin{eqnarray}
\langle\sigma v\rangle_{h_1h_1} &=& \frac{1}{{64\pi m_A^2}}\bigg|g_{A A h_1h_1} + \frac{1}{{D_{h_1}}} g_{h_1A A}  g_{h_1h_1h_1} + \frac{1}{{D_{h_2}}} g_{h_2A A}  g_{h_2h_1h_1} + \frac{{2 g_{h_1A A } ^2}}{{(m_{h_1}^2 - 2 m_A^2)}}\bigg|^2 (1 - m_{h_1}^2/m_A^2)^{1/2} ,\\
\langle\sigma v\rangle_{h_2h_2} &=& \frac{1}{{64\pi m_A^2}}\bigg|g_{A A h_2h_2} + \frac{1}{D_{h_1}} g_{h_1A A } g_{h_1h_2h_2 }+ \frac{1}{D_{h_2}} g_{h_2A A}  g_{h_2h_2h_2} +
\frac{{2 g_{h_2A A} ^2}}{{(m_{h_2}^2 - 2 m_A^2)}}\bigg|^2 (1 - m_{h_2}^2/m_A^2)^{1/2} ,\\
\langle\sigma v\rangle_{h_1h_2} &=& \frac{1}{{32\pi m_A^2}}\bigg|g_{A A h_1h_2} + \frac{1}{D_{h_1}} g_{h_1A A}  g_{h_2h_1h_1} + \frac{1}{{D_{h_2}}} g_{h_2A A}  g_{h_1h_2h_2} + \frac{{g_{h_1A A}  g_{h_2A A }}}{{(m_{h_1}^2 - 2 m_A^2)}}+ \frac{{g_{h_1A A}  g_{h_2A A }}}{{(m_{h_2}^2 - 2 m_A^2)}}\bigg| \nonumber\\
&&\times\sqrt {(1 + \frac{{m_{h_1}^2 - m_{h_2}^2}}{{4m_A^2}}) - \frac{{m_{h_1}^2}}{{m_A^2}}} ,\\
\langle\sigma v\rangle_{h_1h_2bb} &=& \frac{{3m_W^2}}{{\pi g^2}} (\frac{m_b}{v})^2\bigg|\frac{{g_{h_1}(\lambda_{hs}g_{h_1} - \lambda_{s}g_{h_2})}}{{{D_{h1}}}} + \frac{{g_{h_2}(\lambda_{hs}g_{h_2} + \lambda_{s}g_{h_1})}}{{{D_{h_2}}}}\bigg|^2 {(1 - \frac{{m_b^2}}{{M_A^2}})^{\frac{3}{2}}} ,\\
\langle\sigma v\rangle_{h_1h_2tt} &=& \frac{{3m_W^2}}{{\pi g^2}} (\frac{m_t}{v})^2\bigg|\frac{{g_{h_1}(\lambda_{hs}g_{h_1} - \lambda_{s}g_{h_2})}}{{{D_{h_1}}}} + \frac{{g_{h2}(\lambda_{hs}g_{h_2} + \lambda_{s}g_{h_1})}}{{{D_{h_2}}}}\bigg|^2 {(1 - \frac{{m_t^2}}{{m_A^2}})^{\frac{3}{2}}} ,\\
\langle\sigma v\rangle_{h_1h_2WW} &=& \frac{{m_W^4}}{{8\pi m_A^2}}\bigg|\frac{{g_{h_1}(\lambda_{hs}g_{h_1} - \lambda_{s}g_{h_2})}}{{{D_{h_1}}}} + \frac{{g_{h_2}(\lambda_{hs}g_{h_2} + \lambda_{s}g_{h_1})}}{{{D_{h_2}}}}\bigg|^2{(1 - \frac{{m_W^2}}{{m_A^2}})^{\frac{1}{2}}}(2 + {(1 - 2\frac{{m_A^2}}{{m_W^2}})^2})\; ,\\
\langle\sigma v\rangle_{h_1h_2ZZ} &=& \frac{{m_Z^4}}{{16\pi {m_A}^2}}\bigg|\frac{{g_{h_1}(\lambda_{hs}g_{h_1} - \lambda_{s}g_{h_2})}}{{{D_{h_1}}}} + \frac{{g_{h_2}(\lambda_{hs}g_{h_2} + \lambda_{s}g_{h_1})}}{{{D_{h_2}}}}\bigg|^2{(1 - \frac{{m_Z^2}}{{{m_A}^2}})^{\frac{1}{2}}}(2 + {(1 - 2\frac{{m_A^2}}{{m_Z^2}})^2})\;.
\end{eqnarray}\label{eq:ann}
\end{widetext}

\end{document}